\documentclass{article}
\usepackage{arxiv}
\usepackage{indentfirst}
\usepackage[utf8]{inputenc} 
\usepackage[T1]{fontenc}    
\usepackage{hyperref}       
\hypersetup{
    colorlinks=true,
    linkcolor=blue,
    filecolor=magenta,      
    urlcolor=cyan,
}
\usepackage{url}            
\usepackage{booktabs}       
\usepackage{amsfonts}       
\usepackage{nicefrac}       
\usepackage{microtype}      
\usepackage{graphicx}
\usepackage[ruled,vlined]{algorithm2e}
\usepackage{amsmath}
\usepackage{float} 
\usepackage{xcolor,colortbl}
\usepackage{changepage}  
\usepackage[square,numbers]{natbib}

\begin{document}

\renewcommand*{\thefootnote}{\arabic{footnote}}

\title{Biometric Performance as a Function of Gallery Size}

\author{
Lee Friedman\\
Department of Computer Science\\
Texas State University\\
San Marcos, Texas, USA, 78666\\
\texttt{lfriedman10@gmail.com} \\
\and
Hal Stern\\
Department of Statistics\\
University of California - Irvine\\
Irvine, California, USA, 92697\\
\texttt{sternh@uci.edu}\\
\and
Vladyslav Prokopenko\\
Department of Computer Science\\
Texas State University\\
San Marcos, Texas, USA, 78666\\
\texttt{v\_p164@txstate.edu}\\
\and
Shagen Djanian\\
Department of Computer Science\\
Aalborg University\\
Aalborg East, 9220,Denmark\\
\texttt{shagendj@cs.aau.dk}\\
\and
Henry Griffith\\
Department of Computer Science\\
Texas State University\\
San Marcos, Texas, USA, 78666\\
\texttt{h\_g169@txstate.edu}\\
\and
Oleg V Komogortsev\\
Department of Computer Science\\
Texas State University\\
San Marcos, Texas, USA, 78666\\
\texttt{ok11@txstate.edu}\\
}

\maketitle

\begin{abstract}
Many developers of biometric systems start with modest samples before general deployment. But they are interested in how their systems will work with much larger samples. To assist them, we evaluated the effect of gallery size on biometric performance.  Identification rates describe the performance of biometric identification, whereas ROC-based measures describe the performance of biometric authentication (verification). Therefore, we examined how increases in gallery size affected identification rates (i.e., Rank-1 Identification Rate, or Rank-1 IR) and ROC-based measures such as equal error rate (EER).  We studied these phenomena with synthetic data as well as real data from a face recognition study.  It is well known that the Rank-1 IR declines with increasing gallery size.  We have provided further insight into this decline.  We have shown that this relationship is linear in log(Gallery Size).  We have also shown that this decline can be counteracted with the inclusion of additional information (features) for larger gallery sizes.  We have also described the curves which can be used to predict how much additional information is required to stabilize the Rank-1 IR as a function of gallery size.  These equations are also linear in log(gallery size).  We have also shown that the entire ROC curve is not systematically affected by gallery size, and so ROC-based scalar performance metrics such as EER are also stable across gallery size.  Unsurpringingly, as additional uncorrelated features are added to the model, EER decreases.  We were interested in exploring what changes in similarity score distributions might accompany these declines in EERs.  For this, we evaluated the effect of number of features and gallery size on key distribution characteristics (median and IQR) of the genuine and impostor similarity score distributions.  We present evidence that these decreases in EER are driven primarily by decreases in the spread of the impostor similarity score distribution.  
\end{abstract}

\maketitle

\date{September 2019}
\thanks{%
The study was funded by 3 grants to Dr. Komogortsev: (1)~National Science Foundation, CNS-1250718 and CNS-1714623, www.NSF.gov; (2)~National Institute of Standards and Technology, 60NANB15D325, www.NIST.gov; (3)~National Institute of Standards and Technology, 60NANB16D293. Dr. Stern's contributions were partially funded through Cooperative Agreement \#70NANB15H176 between NIST and Iowa State University, which includes activities carried out at Carnegie Mellon University, University of California Irvine, and University of Virginia.
}


\section{Introduction}

Many developers of biometric systems start with modest samples before general deployment. But they are interested in how their systems will work with much larger samples. To assist them, we evaluated the effect of gallery size on biometric performance.  As a general matter, several authors consider that there is an influence of gallery size on biometric performance.  Jain~\cite{JainWeb} has stated: "... that the accuracy estimates of biometric systems are dependent on a number of test conditions, including sensor characteristics, number of subjects in the database,...".  No specific examples are cited in this reference however.   Similarly, Chan et al.~\cite{ChanChallenges} state "...as the number of subjects increases, it becomes increasingly difficult for the system to accurately classify users."  In this case as well, no citations supporting this statement are provided.    

We start with the hypothesis that as gallery size increases, more information is required to achieve any particular level of performance.  We will be addressing performance in terms of identification rate, specifically Rank-1 Identification Rate (Rank-1 IR), and also in terms of ROC-based measures such as EER. (All error rates rates are expressed as percent in the present study.)  It is known that the Rank-1 IR declines with increasing gallery size (see Figure 1).  We will attempt to further describe this relationship, and also show that the decrease in Rank-1 IR can be reversed with the addition of new information. 

\begin{figure}[htbp]
    \centering
    \includegraphics[trim={0.5em 0 1.5em 0}, clip, width=0.85\linewidth]{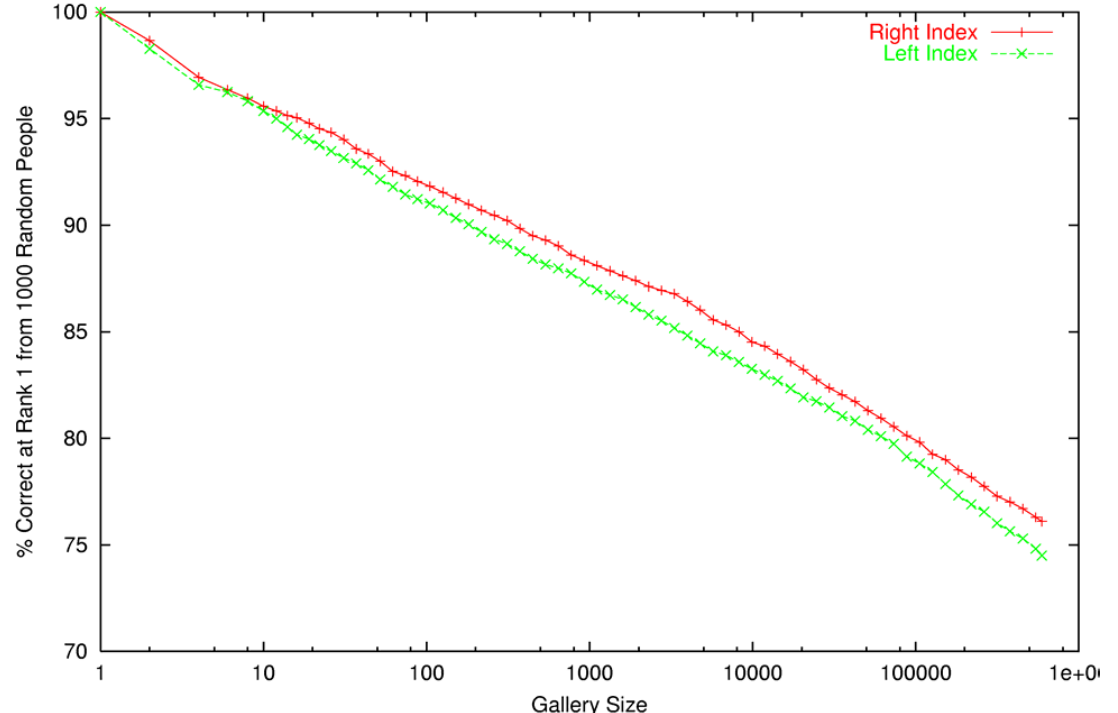}
\end{figure}

There is little published research on the impact of increasing gallery size on ROC-based measures such as EER.  Although it is well established that increasing the number of subjects will decrease the confidence limits on any ROC curves produced~\cite{Jain_CI,Schuckers}, this evidence does not imply anything about the central tendency of any estimated error rate.  

Two earlier papers from our laboratory have addressed the relationship between number of subjects in a database and Rank-1 IR and EER~\cite{OPC,RigasScaling}.  In these studies, biometric authentication was performed using various eye movement features.  The goal was to authenticate subjects tested on two occasions.  The total sample was either 200~\cite{RigasScaling} or 335 subjects~\cite{OPC}. For the purpose of evaluating "Database Scaling Performance", the ROC analysis was repeated for randomly drawn subsets of subjects (\textit{N}=50 to 200 or 335).  For Rank-1 IR, the authors report a
slight reduction in rank-1 identification rates as the subject
pool increases \cite{OPC}. For EER, the authors report that "there was no discernible difference in equal error rates produced for a subject pool of 50 or a subject pool of 323".

We were interested in evaluating the influence of changes in gallery size on a much larger scale (up to 100,000 subjects).  For reasons of convenience, availability and control, we employed synthetic data sets.   However, at each step in the analysis, we provide comparison analyses for a real face-recognition data set (MORPH-II) ~\cite{MORPH} with \textit{N} = 13,930 subjects.  The real face recognition data allowed us to evaluate if the substantive findings with synthetic data were replicated in real biometric data.

In the second section of the manuscript, we present a method for creating synthetic data sets with a number of properties that are helpful for studying biometric performance. Because the data are synthetic, we are able to control the degree of temporal persistence of the features while also ensuring that features are approximately independent of each other and thus provide unique pieces of information for biometric verification. (The concept of "temporal persistence" and the method for its measurement are covered in~\cite{PlosOne}). (In other contexts, this measure is used to assess the inter-rater reliability of a feature.) We think that having unique pieces of information will allow us to address several theoretical notions relevant to biometric analysis in this and subsequent studies. Also in the second section, we present our methods for biometric performance assessment of the synthetic data. In the third section we describe the MORPH-II face recognition data set and our face recognition analyses, including biometric performance assessment.  In the fourth section, we discuss our results for Rank-1 IR. In the fifth section we describe our results for ROC-based metrics. We end with a discussion (section 6).

%

\section{Creation and Analysis of Synthetic data sets}
\label{sec:creation}

\subsection{Creation of Synthetic Data}

\quad (\textit{Please note that these exact procedures for creating synthetic datasets is also described in an manuscript available at (\color{blue}https://digital.library.txstate.edu/handle/\color{black}) currently under review at ACM TOPS.})

Recall that the intraclass correlation coefficient (ICC) is a measure of the correlation expected for repeated measurements of the same feature on different occasions. Unlike the Pearson r correlation coefficient, which is typically applied as an interclass measure of relative agreement (i.e., two series can be correlated even if they differ substantially in level and spread), the ICC is an intraclass measure of absolute agreement~\cite{RN1513}. Measures from the same set of subjects at two different times are intraclass measurements (same metric and variance). ICC ranges from 0.0 to 1.0 with the latter corresponding to perfect temporal persistence. Our goal is to create synthetic features with a specified target ICC (denoted $ICC_{Target}$). Let $X_{ijs}$ denote the measurement of feature $j$ ($j = 1, \dotsc, K$) on session (occasion) $s$ ($s = 1, \dotsc, S$) for individual $i$ ($i = 1, \dotsc, N$). Although the ICC can be calculated based on many sessions, in our experience, biometric assessment is typically performed comparing only two points in time. Therefore, henceforth we will set $S = 2$. We generate normally distributed features such that the theoretical intraclass correlation of repeated measurements of the same feature on the same subject is $ICC_{Target}$ while the theoretical correlation of measurements of different features on the same individual and the theoretical correlation of measurements from different individuals are zero. In practice when data are simulated there are small variations in the empirical ICCs and there are small intercorrelations between features (and individuals) due to chance. 

The algorithm that we use is described briefly here and spelled out in Algorithm~\ref{alg:synthetic}. The starting point is to populate the full set of session one measurements $X_{ij1}$ with random draws from a standard normal distribution (mean zero and variance one). Then the measurements for the second session are set equal to the value of the given feature from the first session,  $X_{ij2} = X_{ij1}$ for ($i = 1, \dotsc, N$, $j=1,\ldots,K$
). At this point both sessions have the same data and each feature has ICC equal to 1.0 (perfect persistence). We obtain the desired ICC by adding a draw from a normal distribution with $\text{mean} = 0$ and $\text{variance} = (1 - ICC_{Target}) / ICC_{Target}$ to each of the measurements. At the end we apply a z-score transform to each feature (with all sessions concatenated together) so that they all have mean 0 and standard deviation one. It can be shown that the resulting measurements have the desired ICC (up to simulation noise). 

\begin{algorithm}
\caption{Creating Synthetic Features}
\SetKwInOut{Input}{Input}\SetKwInOut{Output}{Output}
\Input{$N$ (subjects), $K$ (features), $ICC_{Target}$}
\Output{3-dimensional ($N \times K \times 2$) feature matrix $X_{ijs}$ with desired correlation structure}
\hrule
\smallskip

for $j = 1, \dotsc K$\\
\quad    for $i = 1,\dotsc N$\\
\quad \quad Set $X_{ij1} = Z$ where $Z$ is a random standard normal deviate.\\
\quad \quad Set $X_{ij2} = X_{ij1}$ \\
\hrule
\smallskip
for $j = 1, \dotsc K$\\
\quad    for $i = 1,\dotsc N$\\
\quad \quad for $s = 1,2$\\
\quad \quad \quad Set $X_{ijs} = X_{ijs} + W$; where $W$ is a random normal deviate with mean = 0 and \\
\quad \quad \quad standard deviation = $\sqrt{(1 - ICC_{Target}) / ICC_{Target}}$\\
\hrule
\smallskip
For each feature $j$, treat $X_{ijs}$ as a single vector of length $N \cdot S$ and apply a z-score 
transform \\to ensure mean = 0 and standard deviation = 1
\label{alg:synthetic}
\end{algorithm}

Using this method, we can create features which are normally distributed, that have specified ICCs, with as many subjects and sessions as we desire. These features all have mean~=~0 and SD~=~1. These features are generally independent, but there are some small intercorrelations between features due to chance. To illustrate the approach, we generated data for 10000~subjects, 1000~features and 2~occasions with $ICC_{Target} = 0.7$. Figure 2(A) shows a histogram of the resulting empirical ICCs. Figure 2(B) shows a histogram of the resulting inter-feature correlations.

\begin{figure}[htbp]
    \centering
    \includegraphics[trim={0.5em 0 1.5em 0}, clip, width=0.85\linewidth]{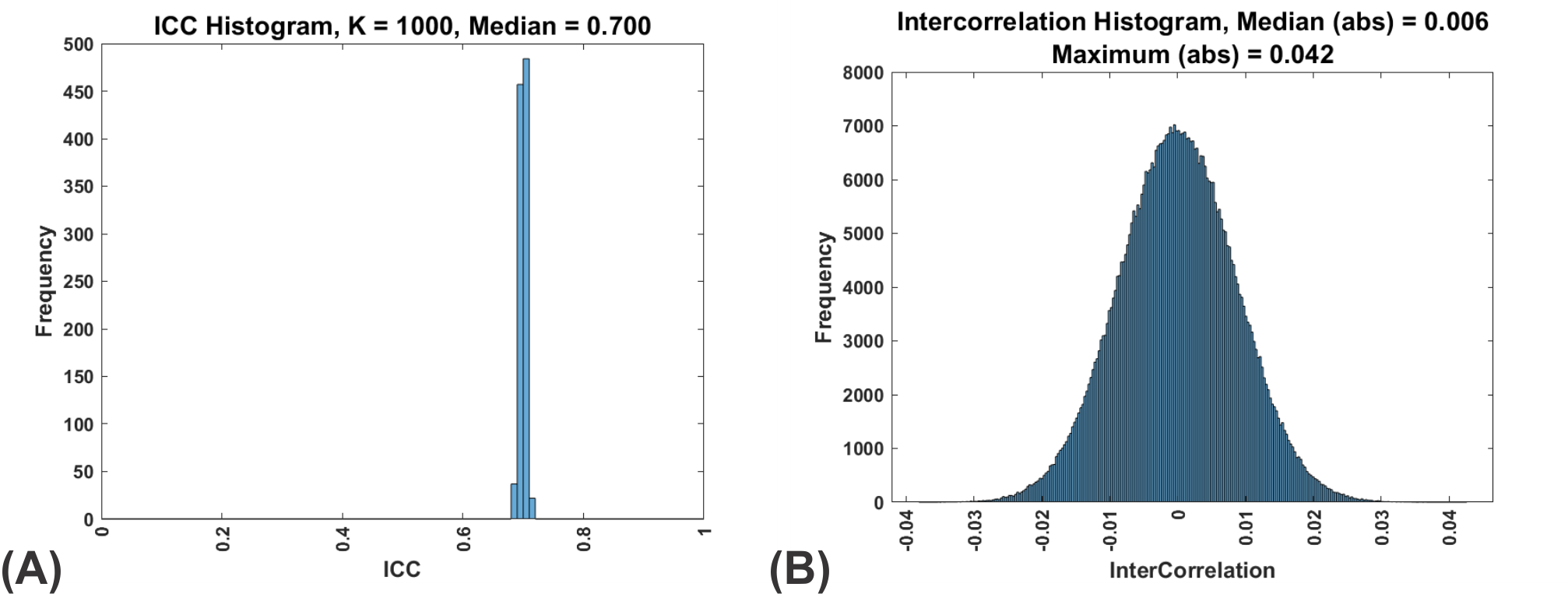}
\caption{{(A)~Frequency histogram of ICCs for 1,000~features with an $ICC_{Target} = 0.7$. This is from a synthetic data set with 10,000~subjects. (B)~Frequency histogram of correlations between 1,000~features for 10,000~subjects, two sessions, with an $ICC_{Target} = 0.7$. Note that the median and maximum are of the absolute value of the correlations.}}
\end{figure}

\subsection{Creation of Sets of Features with Varying Degrees of Persistence}
To match what is found in our face recognition data set (see below), we generated a synthetic data set with varying ICCs in the range of 0.8 to 0.9 (0.8 < ICC <= 0.9). We refer to this data set as a Band 8 data set.  Within the ICC band, the ICCs are evenly distributed across the 0.8 to 0.9 range. 

\subsection{Biometric Performance Assessment for Synthetic Features}
All of the synthetic analyses for this report are based on Band 8 ( 0.8 < ICCs <= 0.9).  This was chosen because in our face recognition data set described below, all of the features had ICCs in this range as well.  A synthetic dataset with 50 features, and 100,000 subjects studied on 2 session with ICC in Band 8 was constructed.  Distance scores were calculated using a variable number of randomly chosen features (2, 5, 8, 10, 15 or 20) from the full set of 50~features. We chose these numbers empirically, for illustrative purposes, based on the range of biometric performance values produced.  For one type of analysis, data set sizes were 1,000, 2,000, 4,000, 8,000, 16000, 32000 or 64,000 subjects.  For another type of analysis, data set sizes were 1,000 (or 2,000), 10,000 or 100,000 subjects. We employed the cosine distance metric, since we have shown in an earlier (unpublished) report that the best biometric performance is produced with this choice\footnote{Link to unpublished report: {\color{blue}https://www.doi.org/10.13140/RG.2.2.17510.06727\color{black}}}. The resulting distance measures were scaled to go from 0~to~1 and then they were reflected ($1 - \text{distance}$) to compute similarity scores. The Rank-1 IR for a data set represents how often (what percent of the time) the greatest similarity score for a probe subject was for the same subject in the gallery. A ``genuine'' distribution of similarity scores was constructed from the similarity scores for each subject and his/her self. All other similarity scores were considered impostors. Applying different decision thresholds to the genuine and imposter similarity scores yields false acceptance and false rejection rates. These can be plotted as a receiver operating characteristic (ROC) curve. The EER is the point on this curve at which the false acceptance rate (FAR) and the false rejection rate (FRR) are equal. 

Since the number of similarity scores is equal to $(gallery \hspace{0.05 in} size)^2$, we can only do exact ROC analyses in computer memory for up to 20,000 subjects.  Therefore we created software which estimates ROC-based measures for much larger sample sizes, and which does not hold all of the similarity scores in memory.  We will call this the large scale procedure for estimating ROC-based measures (LSP-ROC).

The core of this analysis is the computation of the genuine and impostor distribution frequency histograms. All the ROC-based statistics are computed from these two histograms. First, both frequency histograms are initialized to zero. We use 1,000,000 bins of equal size. Cosine similarity scores can assume values from 0 to 1, so each bin has a width of 1/1,000,000. Then both histograms are computed from genuine and impostor similarity scores of subjects from session 1 and session 2.  The similarity scores are calculated iteratively, in batches of 1,000 subjects from each session per batch (1,000,000 similarity scores at a time).

False rejection rate (FRR) and false acceptance rate (FAR) values are computed based on the genuine and impostor distributions frequency histograms: one FRR and FAR value per histogram bin. Let FRR(Similarity score) and FAR(Similarity score) functions be piecewise linear functions based on these FRR and FAR values. EER is computed as a value at the intersection of FAR(Similarity score) and FRR(Similarity score) functions. 

The statistics of genuine and impostor distributions are calculated based on a random sample from the respective \textbf{PDFs} approximated by the relative frequency histograms (calculated from the aforementioned frequency histograms)\footnote{Code is available at \color{blue}https://github.com/v-prokopenko/big\textunderscore roc \color{black}}.

\section{Face Recognition Data and Methods}

\subsection{Data Set and Image Preparation}

We would have preferred to find a publicly available data set with 100,000 or more subjects.  We were not successful.  However, we did find the \textbf{MORPH} Craniofacial Longitudinal Morphological Face Database (\textbf{MORPH-II}) ~\cite{MORPH} \url{(http://www.faceaginggroup.com/morph/)}.  The \textbf{MORPH-II} data set contains mug shots for 13,930 subjects.  Since biometric performance assessment requires at least 2 images per subject, subjects with only one image were excluded (N=857).  The images are colored (RGB), have various dimensions, include more than just the face, and are not spatially registered.  We employed the Viola-Jones algorithm for face detection.  The Viola-Jones algorithm failed in 836 of 53,404 total images.  Viola-Jones algorithm failures accounted for 69 subjects being lost.  This left 13,004 subjects with 2 or more images for further analysis.  Prior to processing these images for facial recognition, the steps in algorithm 2 were applied.

\begin{algorithm}
\caption{\textbf{Steps in the preparation of images for facial recognition}}
\begin{enumerate}
    \item Detect faces in the images using the Viola-Jones algorithm and save the face-only images.
    \item Register the face-only images using an affine transformation (translation, rotation, scale,\\ and shear).
    \item Save registered images as 120 X 100 pixels.
    \item For each subject, correlate each image (as gray scale) with each other image
    \item For each subject, choose the 2 most highly correlated images, and discard all other images.
\end{enumerate}
\end{algorithm}

Note that our goal is not to provide a fair assessment of our face recognition approach to the \textbf{MORPH-II} data set.  Rather, it was to perform a reasonable face recognition analysis which could then be used to evaluate gallery size effects.  For this purpose, we wanted a full range of performance, and our choice to use only the most highly correlated pairs of images was designed to obtain excellent performance under optimal conditions.

\subsection{Face Recognition Approach - FaceNet}

For face recognition, we employed the set of features supplied by the FaceNet algorithm \cite{FaceNet}. 
FaceNet is a deep convolutional network designed by Google, trained to solve face verification, recognition and clustering problems with efficiency at scale.  It is highly accurate and robust to occlusion, blur, illumination, and steering.  It directly maps face images to a compact Euclidean space, where distances directly correspond to a measure of face similarity. Once this space has been produced, tasks such as face recognition can be easily implemented using standard techniques. It achieved accuracy of 99.63\% on the Labeled Faces in the Wild (LFW) data set, and 95.12\% on the YouTube Faces Database.

To create the FaceNet features, we applied the python Keras implementation \footnote{Available at: \color{blue}https://github.com/nyoki-mtl/keras-facenet\color{black}}.  We employed the pretrained Keras model (trained using the Microsoft-Celeb-1M data set (\color{blue}https://megapixels.cc/data sets/msceleb/ \color{black})).
For this, the images needed to be resized to 160x160 pixels and globally rescaled (compute the mean and SD across all intensities from each color channel and, for each channel, subtract the mean intensity and divide by the SD intensity).  The algorithm produced 128 numerical features per subject.

\subsection{Checking Distributions for Normality of FaceNet features}

To assess the normality of the FaceNet features, we computed the skewness and kurtosis of each feature.  The normal distribution has a skewness of 0 and a kurtosis of 3.0.
The 128 FaceNet features had a skewness range of -0.23 to 0.34 and a kurtosis range of 2.76 to 3.33.  On the basis of these ranges, we considered it reasonable to treat all of the FaceNet features as normal.  For comparison, the skewness of a random uniform distribution (k=10,000) was -0.015 and the kurtosis was 1.82.  A log-normal distribution with mean(log(x)) = 0 and sd(log(x)) = 1 had a skewness of 5.35 and a kurtosis of 59.16.

\subsection{Obtaining the ICC and the Feature Intercorrelation for the FaceNet Features}

To further characterize the FaceNet features, we were interested in determining the temporal persistence of the features.  For this we computed the ICC of each feature and present a frequency histogram of these ICCs in Figure 3, left.  All of the features fall between ICC = 0.8 and 0.9.  This means that the features are all highly reliable and are very similar to our synthetic features for Band 8.

\begin{figure}[htbp]
    \includegraphics[trim={0 0 2em 0}, clip, width=1.0\linewidth]{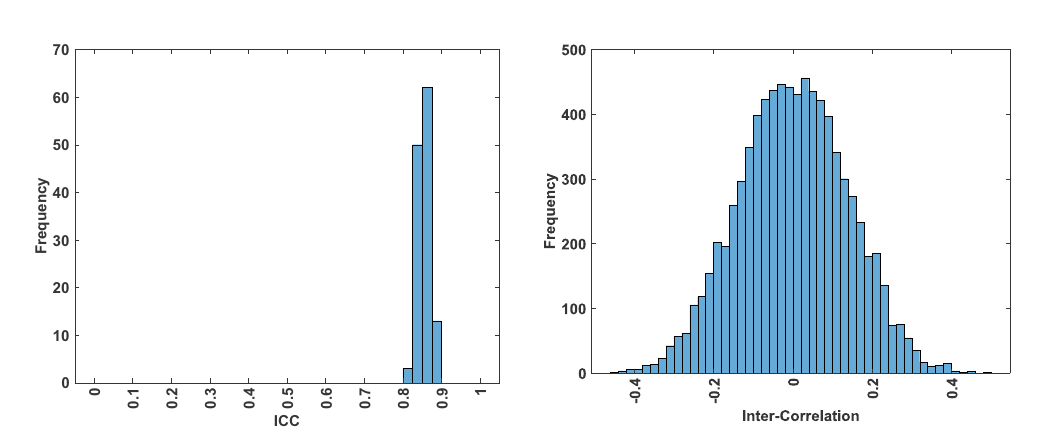}
    \caption{{Characteristics of the FaceNet features.  On the left, we have a frequency histogram of the ICCs for the 128 FaceNet features. All of the features are between 0.8 and 0.9, which corresponds exactly with our synthetic Band 8 features.  This indicates that these features are very reliable over time. On the right we have a frequency histogram of the intercorrelations of the 128 FaceNet features.}}
\end{figure}

The FaceNet features were substanially intercorrelated (Figure 3, right).  For this reason, we decided to perform a PCA on these features.

\subsection{PCA Analysis of FaceNet Features}

In order to create a set of uncorrelated features and to reduce the dimensionality of the data set, we performed a PCA analysis on the FaceNet features.  This analysis included the data for all subjects.  However, during biometric assessment of subsets of the entire dataset (see below), PCA was independently performed on each subset.  As is clear from Figure 4, only approximately 60 uncorrelated features were required to explain 100\% of the variance in the 128 FaceNet features.

\begin{figure}[htbp]
    \includegraphics[trim={0 0 2em 0}, clip, width=0.95\linewidth]{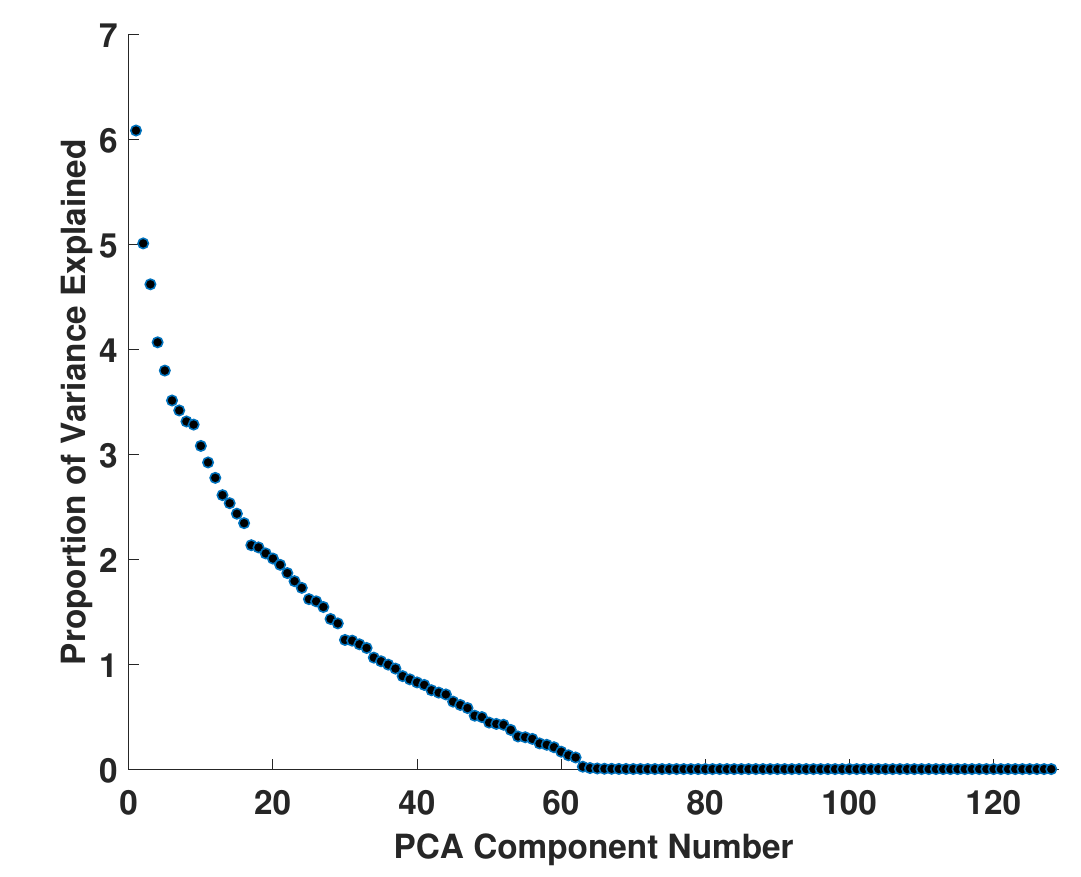}
    \caption{{After PCA of the 128 FaceNet Features, we plot the variance explained by each PCA component against component number. Essentially all of the variance is accounted for by approximately 60 completely uncorrelated PCA components.}}
\end{figure}

\subsection{Biometric Performance Assessment for PCA Components from FaceNet Features}
For one type of analysis, biometric performance was evaluated for the first 2, 5, 12 or 30 FaceNet PCA components. For another type of analysis, biometric performance was evaluated for the the first 5, 10, 15 or 20 PCA components.  Gallery sizes ranged from 1,000 to 10,000 in steps of 1,000.  After a subset of the 13,004 subjects was selected as a data set, PCA was computed on all the first images of each pair.  The PCA coefficients computed from the first images were then used to calculate the PCA components for the second images.  Biometric performance was evaluated in memory (i.e., the LSP-ROC was not needed). Cosine distances were computed, converted to similarity scores, and subjected to a conventional ROC analysis.

\clearpage
\section{Results: Rank-1 Identification Rate}

The Rank-1 IR performance for synthetic Band 8, for 10 features, evaluated at various gallery sizes (1000, 2000, 4000, 8000, 16000, 32000, 64000) is presented in Figure 5 (Top).  Every dot in this figure represents the mean across 30 random repetitions (random subset of features and subjects). The Rank-1 IR for 1,000 subjects was approximately 58\%.  This rate declines steadily as the gallery size increases.  The decrease is described by a linear function of log(Gallery Size).  At a gallery size of 64,000 the Rank-1 IR has dropped below 15\%.

\begin{figure}[htbp]
    \includegraphics[width=0.73\linewidth]{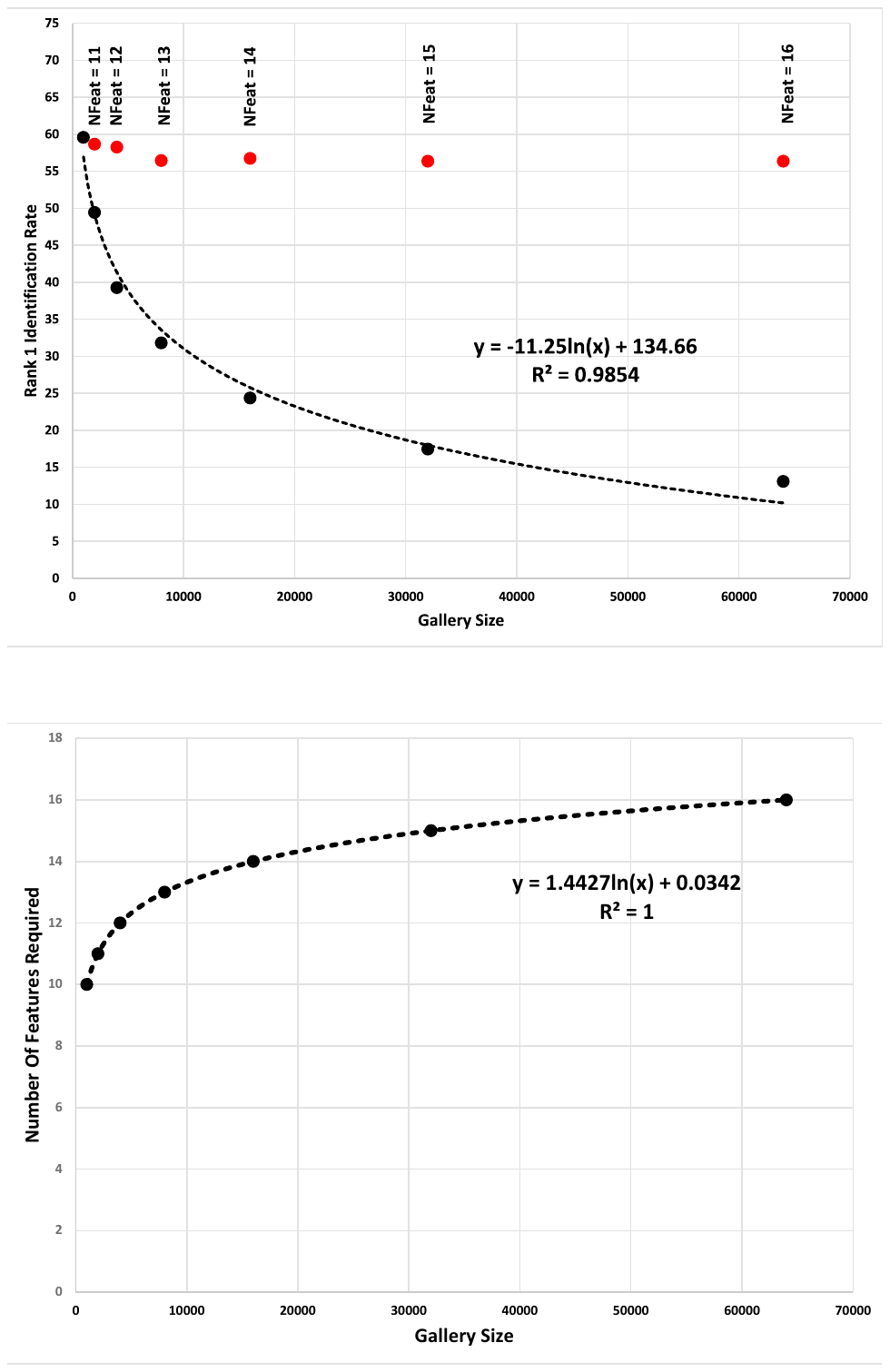}
    \caption{\textbf{Top:} Rank 1 identification rate as a function of gallery size for synthetic Band 8 features. Each dot represents the mean of 30 repetitions. The black dots are the Rank 1 rates for 10 features for the following gallery sizes: 1000, 2000, 4000, 8000, 16000, 32000, 64000 subjects. Note the fit of the decline to a linear function of log(Gallery Size).  The red dots represent the Rank 1 rate for feature numbers greater than 10 which are chosen to produce a Rank-1 rate most similar to that for 10 features. \textbf{Bottom:} Plot of the number of features required for gallery sizes greater than 1000 to match the Rank 1 identification rate for 10 features, 1000 subjects.  In this case a linear equation in log(Gallery Size) is able to match the results perfectly (r-squared = 1.0).}
\end{figure}

\pagebreak \newpage \pagebreak
This drop can be prevented if additional information is added in the form of additional features.  We understand that, in any real world application, investigators would likely use all of the information available in the first instance.  Nonetheless, we believe this additional analysis provides some insight and guidance for improving Rank-1 IR performance.  The red dots represent the Rank-1 IR that can be achieved by adding additional features. The number of features needed to achieve this performance is also indicated.  Note that this analysis is limited in accuracy by the discrete nature of each feature. 

In Figure 5 (Bottom), we plot the number of features required to stabilize the Rank-1 IR as gallery size increases.  In this case, the best fitting function for these feature numbers is also a linear function in terms of log(Gallery Size).

\clearpage
In figure 6 (Top) we present a comparable analysis for our face recognition data set.  In this case the Rank-1 IR is calculated for the first 10 PCA components, as a function of gallery size from 1,000 to 10,000 in steps of 1,000.  Once again, we see a decrease in Rank-1 IR as gallery size increases.  The decrease is linear in log(gallery size).  The number of PCA components required to stabilize the Rank-1 IR performance is also a linear function of log(gallery size)(Figure 6, Bottom).  Unlike the case with synthetic features, each PCA component accounts for distinct amounts of variance in the data.  This analysis treats the components as if they were interchangeable which, formally, they are not.  We think this analysis is nonetheless reasonable on the basis of the fact that the amount of variance accounted for by added components 11 to 14 are not so distinct (2.92, 2.78, 2.61, 2.53).

\begin{figure}[htbp]
     \includegraphics[width=0.65\linewidth]{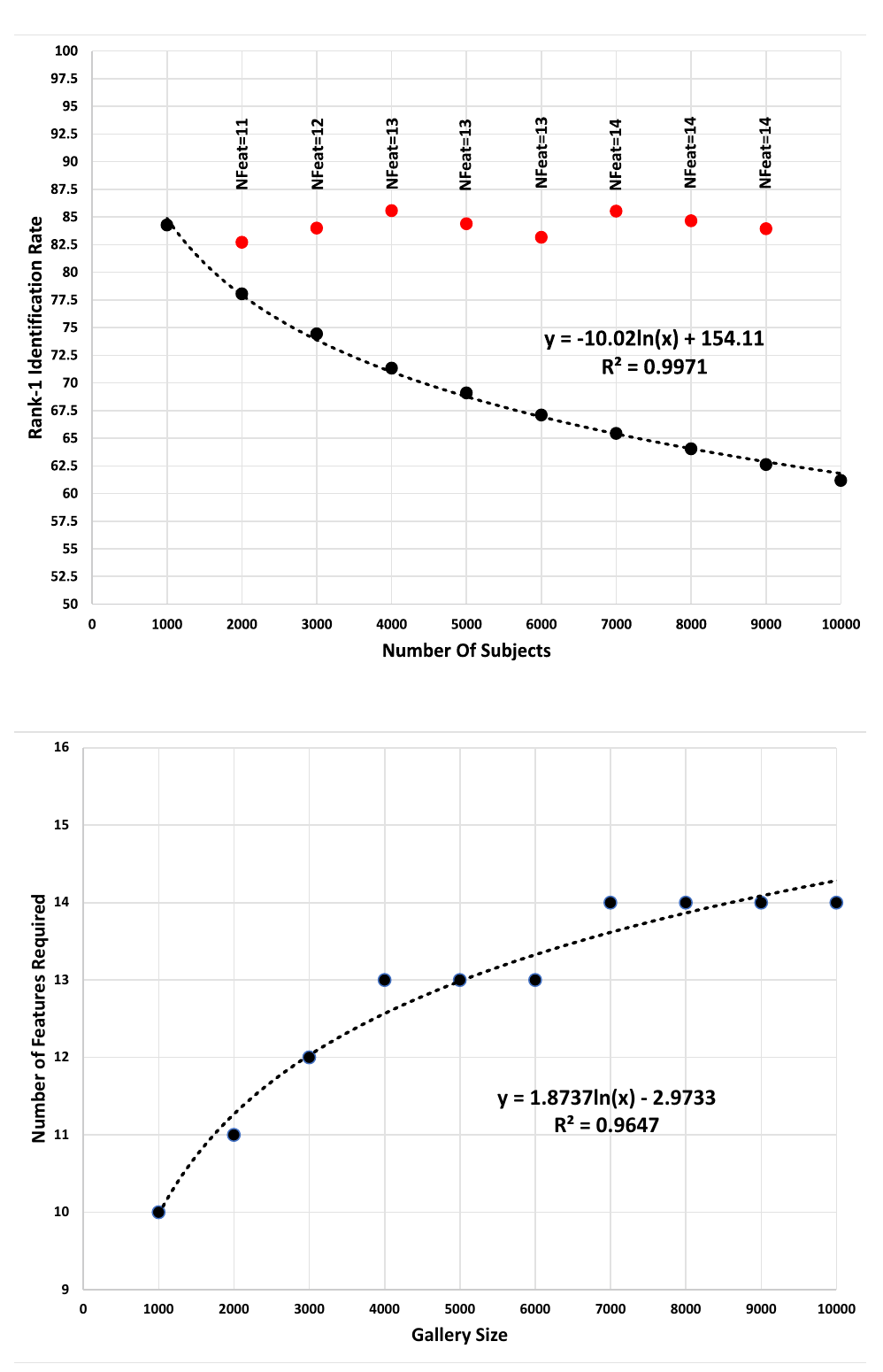}
    \caption{\textbf{TOP:} Rank 1 identification rate as a function of gallery size for FaceNet PCA components. Each dot represents the mean of 30 repetitions. The black dots are the Rank 1 rates for 10 PCA components for gallery sizes from 1000 to 10000 in N=1000 size steps. The red dots represent the Rank 1 rate for feature numbers greater than 10 which are chosen to produce a Rank-1 rate most similar to that for 10 PCA components (1000 subjects). \textbf{Bottom:} Plot of the number of PCA components required for gallery sizes greater than 1000 to match the Rank 1 identification rate for 10 PCA components, 1000 subjects.  A linear function of log(Gallery Size) matches the results quite well (r-squared = 0.96).}
\end{figure}

\section{Results: ROC-Based Measures}

\subsection{Equal Error Rater (EER)}

Figure 7 (Top)  illustrates the EER across gallery size for synthetic data evaluated at 5, 10 15 and 20 features.  Each point is the mean of 30 repetitions (random subset of features and subjects).  Mean EER is apparently very stable across gallery size.

\begin{figure}[htbp]
    \centering
    \includegraphics[width=0.75\textwidth]{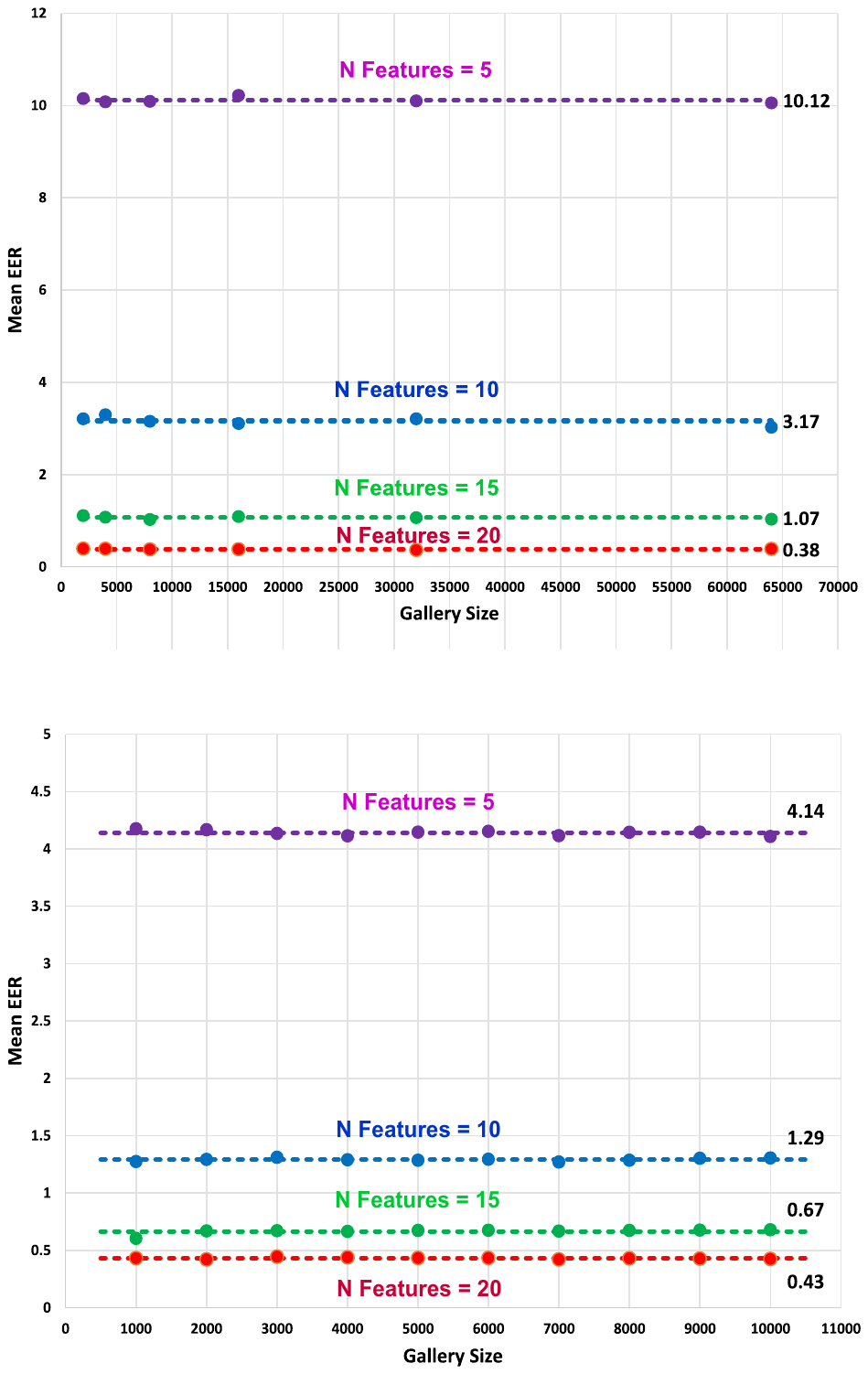}
    \caption{\textbf{TOP:} Relationship between mean EER (30 repetitions) and gallery size for synthetic data (Band 8). Data shown for 5, 10, 15 and 20 features. The average EERs (shown to the right of the last dot) across gallery size are illustrated with dotted lines.  Mean EER is highly stable with increases in gallery size.  \textbf{BOTTOM:} Relationship between mean EER (30 repetitions) and gallery size (1000 to 10000 subjects in 1000 subject steps) for FaceNet PCA components. Data shown for the first 5, 10 15 and 20 PCA components.  The average EERs (shown to the right of the last dot) across gallery size are illustrated with dotted lines. Mean EER does not change as a function of gallery size.}
\end{figure}

Figure 7 (Bottom) illustrates the EER across gallery size for our face recognition data evaluated at the first 5, 10, 15, and 20 PCA components.  Each point is the mean of 30 repetitions (over randomly chosen subjects).  Note the stability across gallery size here.

\clearpage
\subsection{Other Points on the ROC Curve}

Since the EER was so stable across gallery size, we thought it important to check other points on the ROC curve. In Figure 8 (Top), for synthetic data, we present the false rejection rater (FRR) when the false acceptance rate (FAR) = 0.1\% for 5, 10, 15 and 20 features.  Note the stability across gallery size.

\begin{figure}[htbp]
    \centering
    \includegraphics[width=0.75\textwidth]{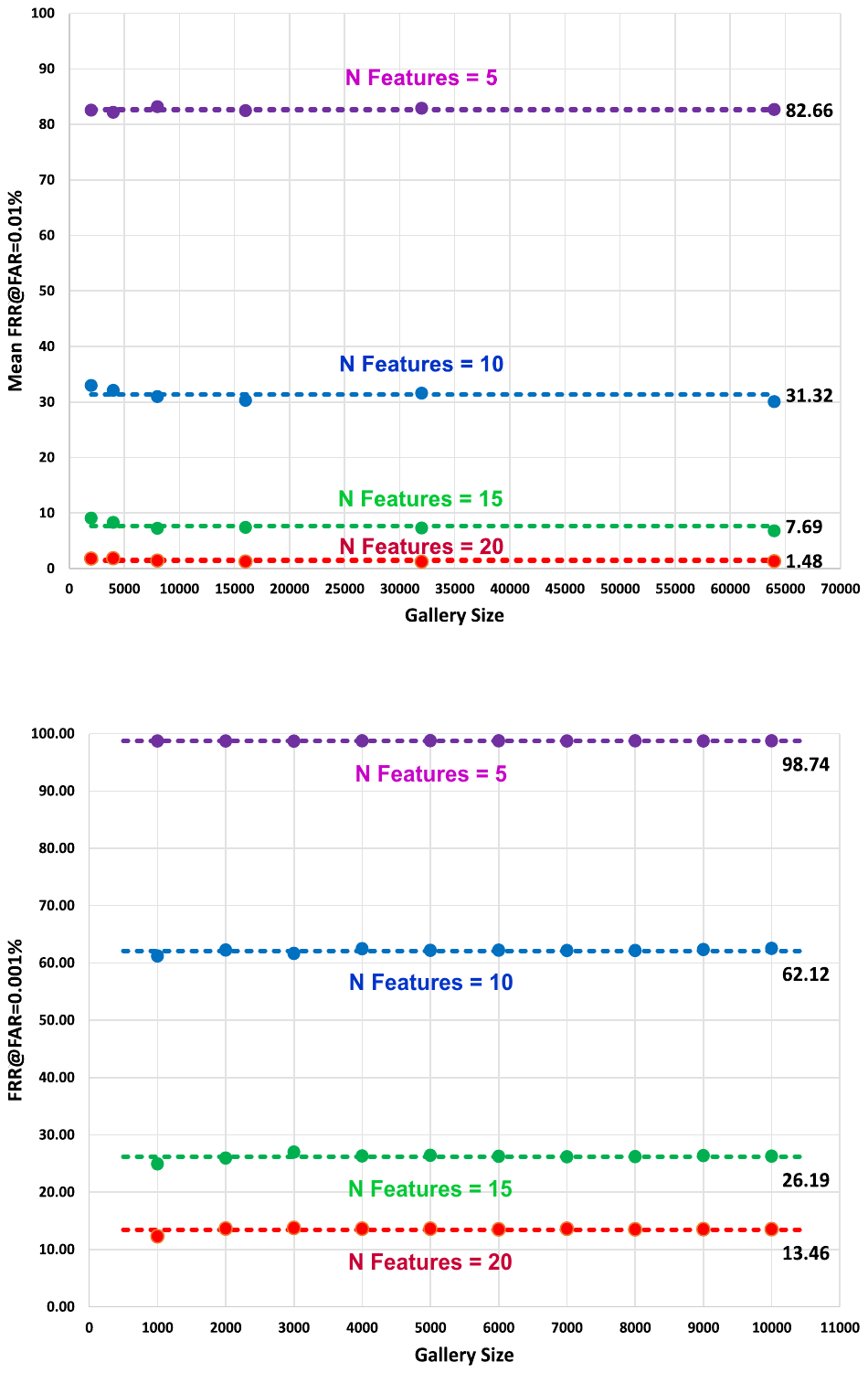}
    \caption{\textbf{Top:}  In addition to mean EER, we also evaluated the mean false rejection rate (FRR) at a false positive rate of 0.1\%. Here we plot these error rates (mean of 30 repetitions) versus gallery size (1000, 2000, 4000, 8000, 16000, 32000, 64000 subjects) for synthetic data (Band 8). Data shown for 5, 10, 15 and 20 features.  The average error rates (shown to the right of the last dot) across gallery size are illustrated with dotted lines. There is some instability at small gallery sizes, but otherwise, the mean error rate does not change as a function of gallery size. \textbf{Bottom:}  FRR at FAR = 0.001\%. Here we plot these error rates (mean of 30 repetitions) versus gallery size (1000 and 10000 subjects) for FaceNet PCA components. Data shown for 5, 10, 15 and 20 PCA components.  The average error rates (shown to the right of the last dot) across gallery size are illustrated with dotted lines.  The mean error rate does not change as a function of gallery size.}
\end{figure}

In Figure 8 (Bottom), for face recognition data, we present similar results for FRR@FAR=0.001\% for our face recognition data set. We analyzed 5, 10, 15, and 20 PCA components. This error rate is quite stable across levels of gallery size.

Neither EER or FRR@FAR=x\% metrics appear to change systematically with gallery size.  Therefore, we hypothesized that the entire ROC curve was also not changing with gallery size.  This is tested below.

\clearpage
\subsection{ROC Curves}

Here we calculate and plot entire ROC-curves for several numbers of features and gallery sizes.  We plot the entire ROC curves (Figures 9) to facilitate the comparison between entire curves as a function of gallery size.

\begin{figure}[htbp]
    \centering
    \includegraphics[width=0.87\textwidth]{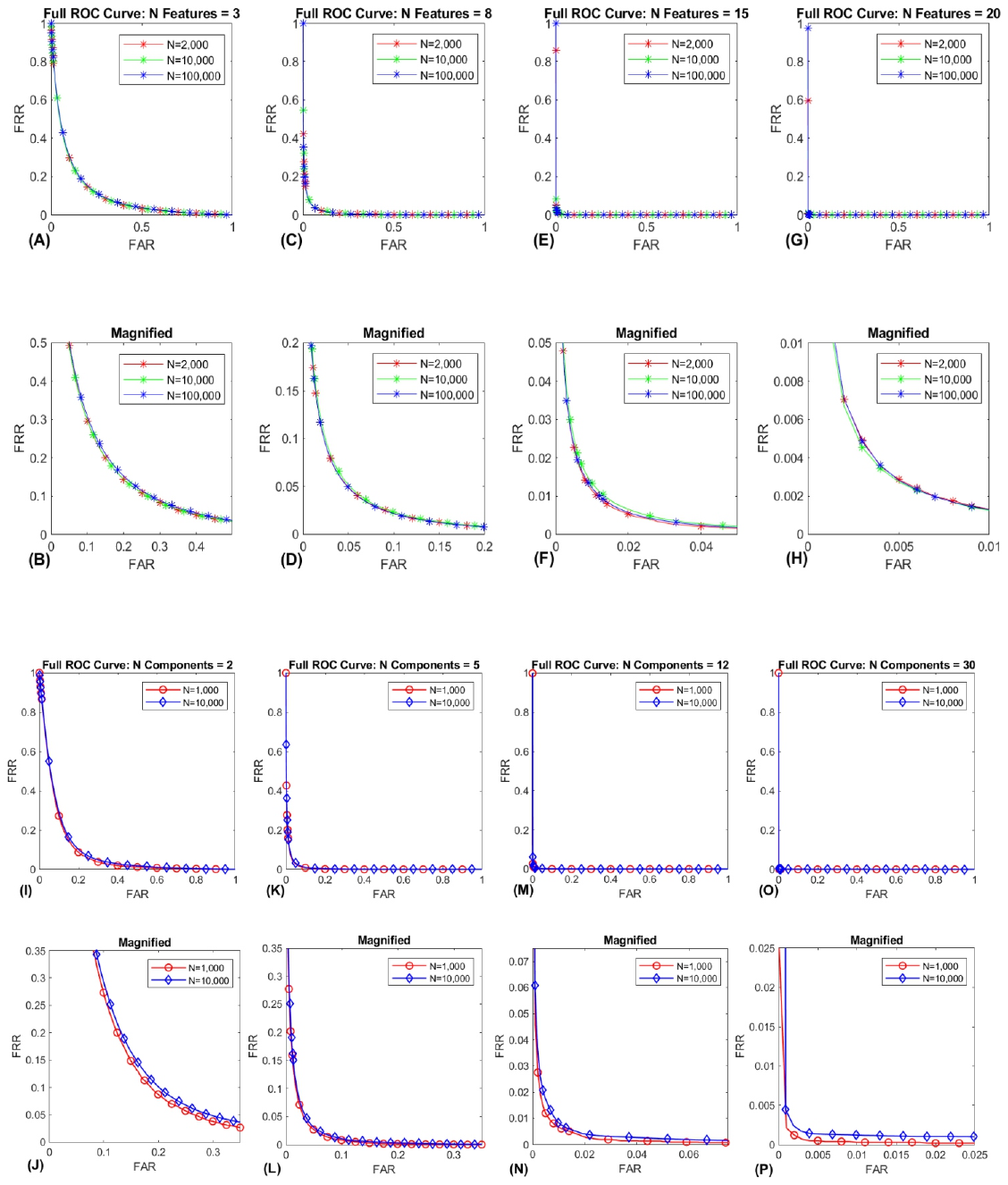}
    \caption{\textbf{Top:} ROC curves for synthetic data (Band 8). Curves for 3, 8, 15 and 20 features.  Each curve is evaluated at gallery sizes of 2,000, 10,000 and 100,000 subjects. Each curve is the average over 30 repetitions.  The top row shows the entire ROC curve (A,C,E,G).  The bottom row is a zoomed in version of the top row, to show the curve at very low error rates (B, D, F, H).  Note that the entire ROC curves are essentially overlapping across gallery size. \textbf{Bottom:} ROC curves for FaceNet PCA components.  Plots for 2, 5, 12 and 30 PCA components are displayed.  Each curve is evaluated at 1,000 and 10,000 subjects.  Each curve is the average over 30 repetitions. The top row shows the entire ROC curve.  The bottom row is a zoomed in version of the top row, to show the curve at low error rates.  Note that the entire ROC curves are essentially overlapping across gallery size.}
\end{figure}

As a general matter, entire ROC curves do not appear to change as a function of gallery size. Since the ROC-curves are based on the characteristics of the genuine and similarity score distributions, it makes sense that the ROC-curves don't change.  With increases in gallery size, the distributions will become more well defined, and the median and IQR should become increasingly stable.  But there is no basis for predicting a change in the central tendency or spread these distributions as gallery size increases.

\clearpage
\subsection{Similarity Score Distribution Metrics}

ROC curves are based on genuine and impostor similarity score distributions. We were interested in evaluating changes in similarity score distributions as a function of gallery size. To this end, we evaluated the median and IQR of both genuine and impostor distributions as a function of number of features and gallery size.  The results are presented in Figure 10.

\begin{figure}[htbp]
    \centering
    \includegraphics[width=0.9\textwidth]{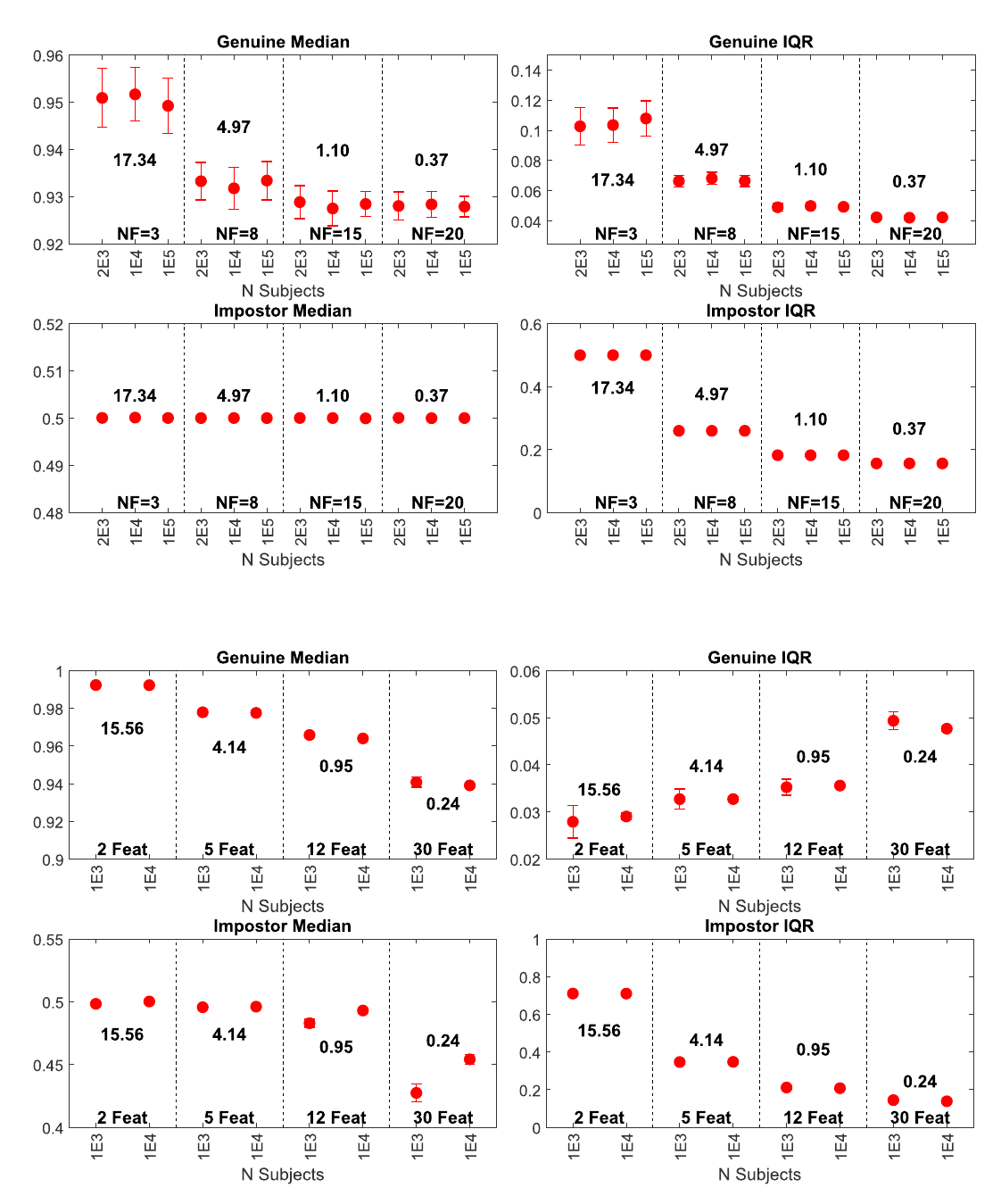}
    \caption{\textbf{Top:} Similarity score distributions characteristics for synthetic data.  Data for the Band 8 data set with 3, 8, 15 and 20 features.  Each dot is based on 30 repetitions. The error bars are at +/- 1 SD. The top row represents the metrics (median, IQR) for genuine distributions. The bottom row represents the same metrics for impostor distributions. The numbers in each plot are the average EER across the gallery sizes represented.  \textbf{Bottom:} Similarity score distribution characteristics for FaceNet PCA components. Data for 2, 5, 12 and 30 PCA components.  Each boxplot is based on 30 repetitions. The top row represents the metrics (median, IQR) for genuine distributions. The bottom row represents the same metrics for impostor distributions. The numbers in each plot are the average EER across the gallery sizes represented.}
\end{figure}

For both Figure 10 (Top and Bottom), for any particular number of features (or PCA components), the metrics appear reasonably stable across gallery size.
If one takes into account the ranges of the plots, the biggest systematic change as a function of number of features is in the impostor IQR.  As new features are added, the impostor distributions are decreasing in spread.  It appears that this effect is driving performance as assessed with EER.  To compare the contribution of each distribution metric (median and IQR) to EER, we performed a stepwise linear regression.  The dependent variable was the mean EER for all twelve levels represented in each plot.   The independent variables were the means of the distribution metrics.  We did this analysis for both synthetic and face recognition data sets.  For both data sets,  the first independent variable entered into the model was for the impostor IQR (synthetic: p = 2.e-15, $r^2$ = 0.998; real: p = 1.8e-6,  $r^2$ = 0.979).  This is consistent with the idea that the decrease in impostor IQR with increasing features is the driving force behind the lower EERs obtained.

The medians for the genuine distributions are steadily decreasing which, all other things held constant, should result in worsening performance.  However the magnitude of these changes is very small.  The IQR of the genuine distributions are decreasing, which should lead to improved performance, but the range of these changes is also small. For synthetic data, the median of the impostor distributions are all exactly 0.5.  For our face recognition data, for higher numbers of features, there is some evidence of a decrease in the impostor distribution median. Note that this particular pattern may be related to the distance metric (cosine) that we have employed, and other aspects of the analysis.  In the future, we hope to evaluate the role of distance metric and other design elements on the pattern of these similarity score medians and IQRs as performance improves.

\section{Discussion}
In this report, we have replicated and extended findings regarding the effects of gallery size on biometric performance.  We have confirmed the finding that   Rank-1 IR declines as a function of gallery size.  We have shown that this relationship is linear in log(Gallery Size).  We have also shown that this decline can be counteracted with the inclusion of additional information (features) for larger gallery sizes.  We have also described the curves which can be used to predict how much additional information is required to stabilize the Rank-1 IR as a function of gallery size.  These equations are also linear in log(gallery size).  

It is important to note that our findings are based on datasets with certain characteristics. In particular we are using either features or PCA components that are reasonably normally distributed. Also, we are using features or PCA components that are either completely uncorrelated or nearly completely uncorrelated.  Any data set with approximately normal features that are more strongly intercorrelated can be transformed into a completely uncorrelated data set using the inverse Cholesky transformation \cite{RN1672}\footnote{See ``Decorrelation of n random variables'' at https://en.wikipedia.org/wiki/Pearson\_correlation\_coefficient}. 

We have also shown that ROC curves are not systematically affected by gallery size, and so ROC-based scalar performance metrics such as EER are also stable across gallery size.  We have illustrated how changes in similarity score distribution characteristics (median and IQR) change as additional features are added to the analysis.  The most important predictor of change in EER as additional features are added is in the spread of the impostor similarity score distribution, which becomes narrower as additional features are added.  This particular pattern of changes may be related to the distance metric chosen (cosine) or other aspects of the design.  The topic of changes in similarity score distributions that accompany changes in biometric performance will be addressed in future work.  These findings should be of interest in the abstract theoretical sense, and should be of real practical value to the biometric community, when planning biometric modalities of various gallery sizes.

The ROC-based measures are based on the central tendency and spread of the genuine and similarity score distributions. Although measures of central tendency and spread will become more stable with increasing gallery size, there is no basis for predicting a systematic change in the central tendency or spread of these distributions with increasing gallery size.

\section{Acknowledgments}
This work was supported by grants to Dr. Komogortsev (NSF \#CNS-1250718, NSF \#CNS-1714623, and NIST \#60NANB16D293).
We wish to acknowledge the assistance of Dillon  J Lohr, a doctoral student in our group.  He was immensely helpful in preparing the \LaTeX\ version of the paper.

\bibliographystyle{spbasic.bst}
\bibliography{MyText.bib}

\end{document}